\documentclass{spie}  
\usepackage[]{graphicx}
\newcommand{\degr}{\ensuremath{^{\circ}}}
\newcommand{\degrC}{\ensuremath{^{\circ}}C\ }

\title{Detecting the full polarisation state of terahertz transients}

\author{E.~Castro-Camus,\supit{a} J.~Lloyd-Hughes,\supit{a} M.D.~Fraser,\supit{b} H.H.~Tan,\supit{b} C.~Jagadish\supit{b} and M.B.~Johnston\supit{a}
\skiplinehalf \supit{a}University of Oxford, Department of Physics, Clarendon Laboratory, Parks Road, Oxford, OX1 3PU, United Kingdom; \\
\supit{b}Department of Electronic Materials Engineering, Research School of Physical
Sciences and Engineering, Institute of Advanced Studies, Australian National University,
Canberra ACT 0200, Australia}

\authorinfo{Further author information: \\
http://www-THz.physics.ox.ac.uk
\\E.C.C.: E-mail: e.castro-camus@physics.ox.ac.uk, Telephone: +44 (0)1865 272339\\  M.B.J.: E-mail: m.johnston@physics.ox.ac.uk, Telephone: +44 (0)1865 272236}

\begin{document}
\maketitle

\vspace{-6.0cm}

\parbox{17.2cm}{\centering{\footnotesize SPIE Photonics West 2006 Invited Talk [6120-29]}}

\vspace{6.0cm}
\begin{abstract}

We have developed a detector which records the full polarisation
state of a terahertz (THz) pulse propagating in free space. The
three-electrode photoconductive receiver simultaneously records the
electric field of an electromagnetic pulse in two orthogonal
directions as a function of time. A prototype device fabricated on
Fe$^{+}$ ion implanted InP exhibited a cross polarised extinction
ratio better than 390:1. The design and optimisation of this device
are discussed along with its significance for the development of new
forms of polarisation sensitive time domain spectroscopy, including
THz circular dichroism spectroscopy.

\end{abstract}


\section{INTRODUCTION}
\label{s:introduction}  
Many collective processes in condensed matter physics and
macromolecular chemistry have energies that fall in the
far-infrared, or terahertz (THz), region of the electromagnetic
spectrum. However, in the past this spectral region has been
relatively unexplored owing to a lack of bright radiation sources
and appropriate detectors. The advances in ultra-fast laser
technology over the last decades allowed establishing the technique
of terahertz time domain spectroscopy (THz-TDS)
\cite{AustonC85,SmithAN88} to perform linear spectroscopy in the far
infrared region of the spectrum. TDS provided a powerful in
condensed matter physics
\cite{HuberTBBAL01,LeitenstorferHTBBA02,KaindlHCLC03} and
macromolecular chemistry.\cite{cpl03,schmuttenmaer2004} This
technique has become popular because of its high sensitivity over a
wide spectral band between 100\,GHz and 10\,THz. In general THz-TDS
is performed using emitters and detectors of linearly polarised
radiation. However, in order study many systems of scientific
interest such as birefringent and optically active materials
properly it is necessary to record the full polarisation state of
the radiation.

A novel technique called vibrational circular dichroism (VCD) has
substantial potential in the fields of macromolecular
chemistry,\cite{salzman:2175} and structural biology.\cite{nafie96}
In addition to the established technique of (ultraviolet) circular
dichroism, VCD is used to analyse the structure of chiral molecules.
It is predicted that VCD will be more powerful than conventional
circular dichroism for stereo-chemical structure
determination.\cite{nafie96} However the technique is currently
limited by insensitive and narrow band spectrometers.

The particular three-dimensional structure assumed by some
biomolecules, marks the difference between an inert chain of amino
acids or monomers and a fully functional protein or nucleic
acid.\cite{usheva:thzp2005} In many of these cases the
three-dimensional structure is chiral causing radiation to interact
differently with them depending on its circular polarisation state
and therefore they are expected exhibit circular
dichroism.\cite{xu19,XuRGSSBAP03} Terahertz radiation falls in the
appropriate energy range of the spectrum to excite vibrational modes
in many systems including biomolecules, the study of circular
dichroism of such collective modes can provide fundamental
information about the structure and strength of inter atomic bonds
in these molecules which is of major interest to biochemists.

\section{Photoconductive detection}
\label{s:pcd}

Terahertz TDS relies on the ability of detecting the amplitude of
the electric field as function of time rather than the intensity of
the radiation, one of the two main methods to measure the electric
field is photoconductive detection. In order to measure a THz
electric field $\mathbf{E}_{\rm THz}$ using a photoconductive
receiver it is necessary to gate the receiver with an ultra-short
laser pulse.  The laser pulse generates free carriers in the
semiconductor changing its conductivity. These carriers are
accelerated by $\mathbf{E}_{\rm THz}$ from the moment of the arrival
of the pulse onwards thus generating a current $I$ between two
contacts, if ${\mathbf{E}}_{\rm THz}$ is delayed by $t'$ with
respect to the laser pulse the current will be given by:
\begin{equation}
{ I(t) \propto \int_{-\infty}^\infty \mathbf{E}_{\rm THz}(t') \cdot
\mathbf{\hat{u}}\, \Phi(t'-t)\rm{d}t' } \label{e:I}
\end{equation}
where $\mathbf{\hat{u}}$ is a unit vector pointing between the
contacts, and $\Phi(t'-t)$ is the response of the semiconductor. If
the laser pulse is approximated by a Dirac delta function and the
recombination and trapping times of the substrate are assumed to be
infinite, the function $\Phi(t'-t)$ will be a step function and the
electric field could be recovered by differentiating the previous
equation. However the laser pulse has a non-zero duration, and the
trapping and recombination time constants of the substrate are
finite. Assuming a laser pulse of the form $\rm{sech}^2(1.76t/t_0)$
where $t_0$ is the full-width-at-half-maximum, and an exponential
carrier trapping with time constant $\tau$ the response of the
substrate would be of the form
$\Phi(t'-t)=e^{-(t'-t)/\tau}[1+\tanh(1.76(t'-t)/
t_0)]$.\cite{kono:898} In this case Equation~\ref{e:I} has a
solution given by:
\begin{equation}
{ E(t) \propto
\mathcal{F}^{-1}\left[\frac{\mathcal{F}[I(t)](\omega)}{\mathcal{F}[\Phi(t)](\omega)}\right](t)
} \label{e:E}
\end{equation}
where $\mathcal{F}$ is the Fourier transform operator. In order to
recover the electric field it is a common practice approximate
$\Phi(t'-t)$ as a step function and differentiate the measured
current, this is a good approximation as long as the laser pulse
remains much shorter than the THz pulse ($<$100\,fs) and the
trapping time remains longer than the duration of the THz pulse
($>$1\,ps). Trapping times much longer than the duration of the THz
pulse ($\gg10$\,ps) are not desirable because the receiver would
stay conductive for a long period of time accumulating noise on top
of the signal.

\section{Polarisation sensitive detector}
\label{s:detector}

\begin{figure}[bt]
    \centering
    \includegraphics[height=5cm]{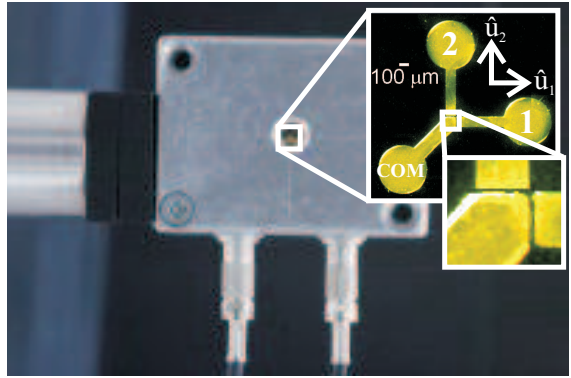}
    \caption{\label{FIG:rec}
    Photograph of the three-electrode photoconductive receiver,
    the insets show microscope photographs of the contact structure.
    This receiver was designed in such a way that the current flowing through gaps formed
    between contacts 1 and the common contact (\textsc{com}) and between contacts 2 and \textsc{com} are
    mutually orthogonal.
    }
\end{figure}

Traditionally photoconductive detectors consist of a pair of
contacts on a semiconductor, typically low temperature grown GaAs or
InP, separated by a small ($\sim 16\,\rm{\mu m}$) gap. The
possibility of building similar detector capable of recording both
components of the electric field of a THz pulse was explored
recently.\cite{castro-camus:254102} This novel type of detectors
consist of three electrodes arranged in the geometry shown in
Figure~(\ref{FIG:rec}). The geometrical structure of this device was
designed in such a way that the vectors $\bf{\hat{u}}_1$ and
$\bf{\hat{u}}_2$ crossing ``gap 1'' (formed between contacts 1 and
\textsc{com}) and ``gap 2'' (between contact 2 and \textsc{com}) are
mutually orthogonal. The current $I_1$ and $I_2$ through the gaps can
be measured simultaneously, replacing these currents in
Equation~(\ref{e:I}) with the respective vector $\bf{\hat{u}}_i$ both
transverse components of the electric field can be calculated.

This method can measure not only the amplitude of both electric
field components but also their relative phase. This allows
resolving the full polarisation state of the THz wave and therefore
measuring complex polarisation dependent dielectric properties of
materials over a broad region of the spectrum in a single
experiment.

\begin{figure}[bt]
    \centering
    \includegraphics[height=7cm]{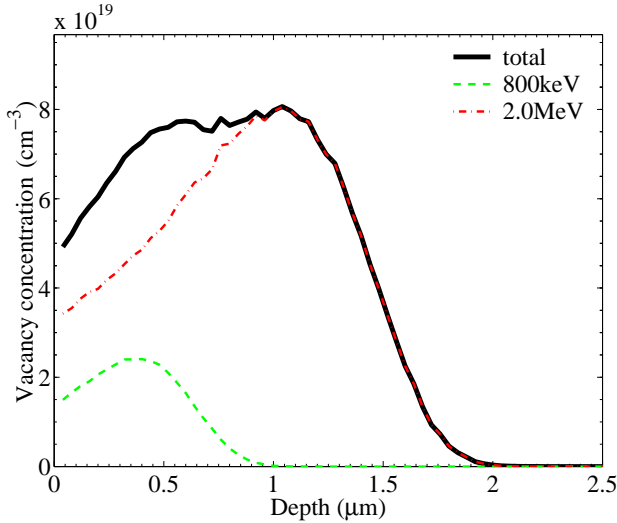}
    \caption{\label{FIG:doses}
    Vacancy distribution profile in InP:Fe$^+$, the InP samples
    were implanted with ions of two different energies in order
    to create an approximately constant vacancy distribution over
    absorption depth ($\sim1$\,$\rm{\mu m}$) of near infrared (800\,nm) light.
    }
\end{figure}

In order to fabricate the three-electrode receiver, a sample of InP
was implanted with Fe$^+$. The implantation was done at two
different energies, $1.0\times10^{13}$\,cm$^{-2}$ at 2.0\,MeV and
$2.5\times10^{12}\,$cm$^{-2}$ at 0.8\,MeV. With this implantation
conditions an approximately constant profile of vacancies was
obtained, as shown in Figure~\ref{FIG:doses}, over the laser wavelength
(800\,nm) absorption depth ($\sim1\,\rm{\mu m}$). The sample was
annealed at 500\degrC for 30\,min under a PH$_3$ atmosphere. With
these implantation and annealing conditions a substrate with a
controlled carrier trapping time constant was produced. The trapping
time was measured by optical pump THz probe spectroscopy to be
$5.9\pm0.7$\,ps.

The InP:Fe$^+$ sample surface was cleaned and spin coated at
6000\,rpm for 60\,s with  photoresist (\emph{Shipley}), producing a
photosensitive film of $\sim 1\,\rm{\mu m}$ thickness. The sample
was then soaked in chlorobenzene for 120\,s and subsequently exposed
and developed. A 20\,nm thick layer of Cr was evaporated on the
sample followed by 250\,nm of Au. The sample was left in acetone
until the Cr/Au layer was lifted leaving only the contacts on the
semiconductors surface as shown in Figure~\ref{FIG:rec}. The sample was
mounted on a DIP-8 chip package and connections to the electrodes
were made using an ultrasonic wire bonder.

\begin{figure}[bt]
    \centering
    \includegraphics[height=7cm]{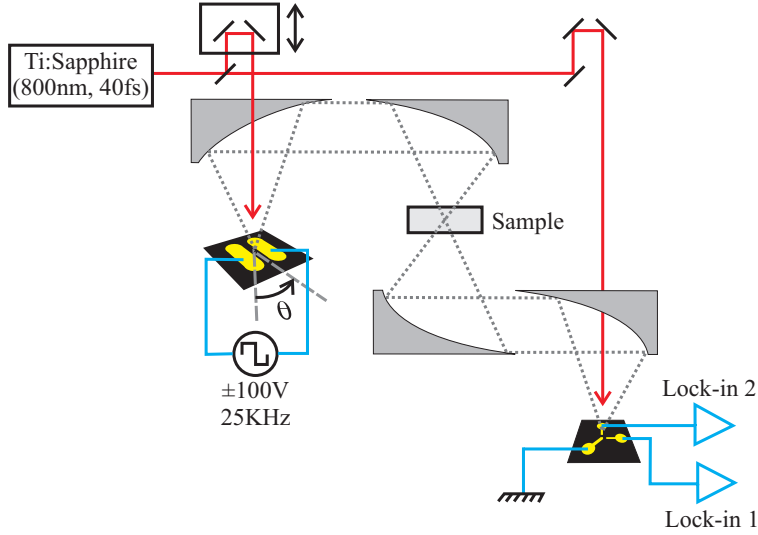}
    \caption{\label{FIG:setup}
    Diagram of experimental apparatus used
for simultaneous detection of horizontal and vertical components of
the electric field of a THz transient. A SI-GaAs photoconductive
switch was used as emitter, and parabolic mirrors were used to collect
and focus the THz radiation onto the three-contact photoconductive
receiver.
    }
\end{figure}

The receiver was tested using the experimental configuration shown
in Figure~\ref{FIG:setup}. A Ti:sapphire oscillator provided 40\,fs,
5.5\,nJ pulses at 80\,MHz repetition rate with a centre wavelength
of 800\,nm. Each pulse was divided using a 90:10 beam splitter.
Linearly polarised THz transients were produced by exciting a
400\,$\rm{\mu m}$ gap semi-insulating GaAs photoconductive switch
with the 90\% portion of the laser pulse. The photoconductive switch
was biased with a $\pm$120\,V square wave at a frequency of 25\,kHz
and was mounted on a rotational stage to control the angle of the
plane of polarisation of the radiation. Using off-axis parabolic
mirrors the THz radiation was collected, focused onto a sample
space, collected again and focused onto the three-contact
photoconductive receiver. The three contact receiver is gated by the
remaining 10\% of the laser pulse.  The current between the
\textsc{com} electrode and electrodes 1 and 2 were amplified by
lock-in amplifiers locked to the 25\,kHz square wave. Both signals
where recorded simultaneously by a multi channel analogue to digital
converter. The THz radiation path was enclosed in a vacuum chamber
at a pressure lower than 1\,mbar to avoid signal attenuation
resulting from water vapor absorption.

\begin{figure}[bt]
    \centering
    \includegraphics[height=7cm]{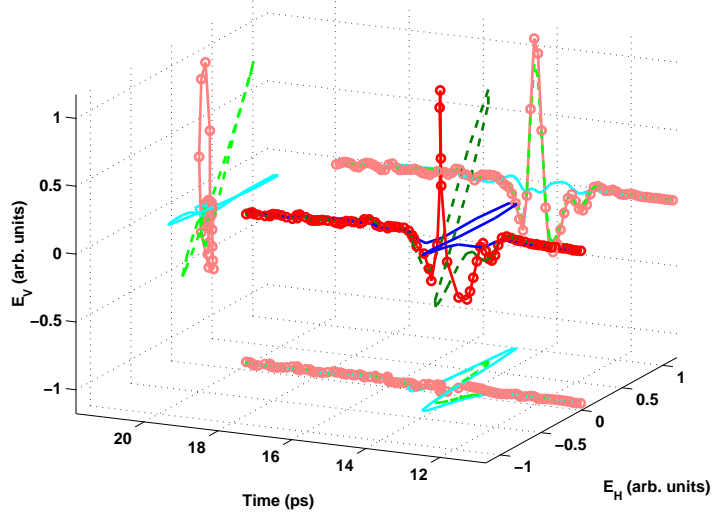}
    \caption{\label{FIG:treeangles}
    Three dimensional representation of electric fields measured at 0\degr, 45\degr and 90\degr.
    }
\end{figure}

By varying the position of the translational stage (see
Figure~\ref{FIG:setup}) the terahertz and probe pulses could be
delayed with respect to each other. To prove the effectiveness of
the polarisation sensitive receiver scans with the emitter at angles
0\degr, 45\degr and 90\degr with respect to the horizontal were
taken. A three dimensional representation of these pulses is shown
in Figure~\ref{FIG:treeangles}. It can be noticed that the three
curves are nearly plane, the angles measured for the signals are
-3\degr, 49\degr and 92\degr, they are all within $\pm5\degr$ of the
expected angle, this is under the uncertainty of the rotational
stage holding the emitter. The crossed polarisation extinction ratio
can give a measure of the effectiveness of the receiver as
polarisation sensitive detector, this ratio was determined to be
better than 390:1.

\section{Characterisation of quartz}
\label{s:quartz}

\begin{figure}[bt]
    \centering
    \includegraphics[height=5cm]{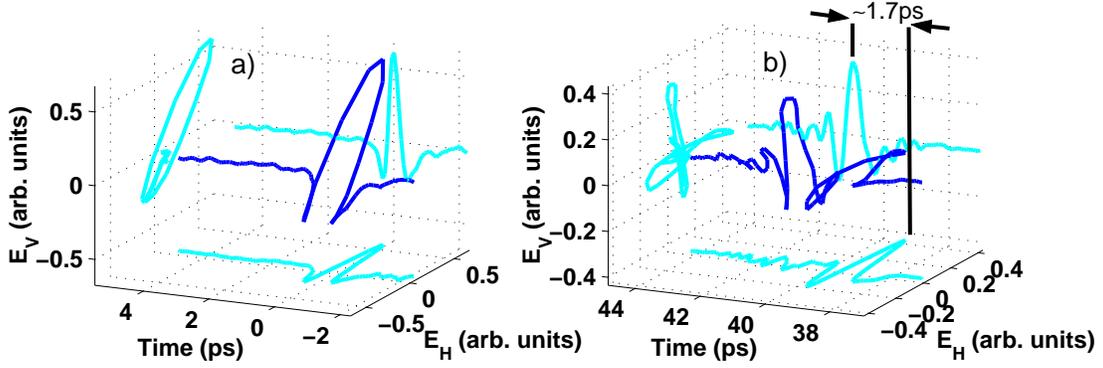}
    \caption{\label{FIG:rnq}
    a) Three dimensional representation of the reference
    electric field as function of time, the curves in gray are the projections on
each of the planes. b) Analogous plot for the electric field through
a 10.54mm sample of x-cut quartz.
    }
\end{figure}

The possibility of measuring both components of the THz wave
considerably extends the power of traditional time domain
spectroscopy. To demonstrate the potential of this extended
technique, we decided to measure the birefringence of quartz that is
a model system among the uniaxial crystals that present this
phenomenon. The system shown in Figure~\ref{FIG:setup} was used. The
emitter was aligned at an angle of 45\degr. A 10.54\,mm thick X-cut
quartz crystal was placed in such a way that its Y and Z axes were
aligned with the horizontal and vertical directions respectively.
The reference electric field ${\bf{E}}_{\rm ref}$ was measured
(Figure~\ref{FIG:rnq}a) before mounting the sample, then the
electric field ${\bf{E}}_{\rm sam}$ with the quartz sample in the
THz path was also measured (Figure~\ref{FIG:rnq}b). Substantial
differences can be noticed between ${\bf{E}}_{\rm ref}$ and
${\bf{E}}_{\rm sam}$. There is a delay of $\sim$40\,ps between the
two waves due to the delay caused by the optical density of the
quartz sample. Also the total amplitude of the wave is considerably
smaller, caused by the Fresnel transmission at the two faces of the
quartz. Additionally an important difference in the shape of the
wave may be observed, the original transient became split into two
linearly polarised transients as it propagated through the medium,
one with the electric field parallel to the optical axis (Z) and one
perpendicular. These ordinary and extraordinary components are
separated by $\sim$1.7\,ps from each other.

\begin{figure}[bt]
    \centering
    \includegraphics[height=5cm]{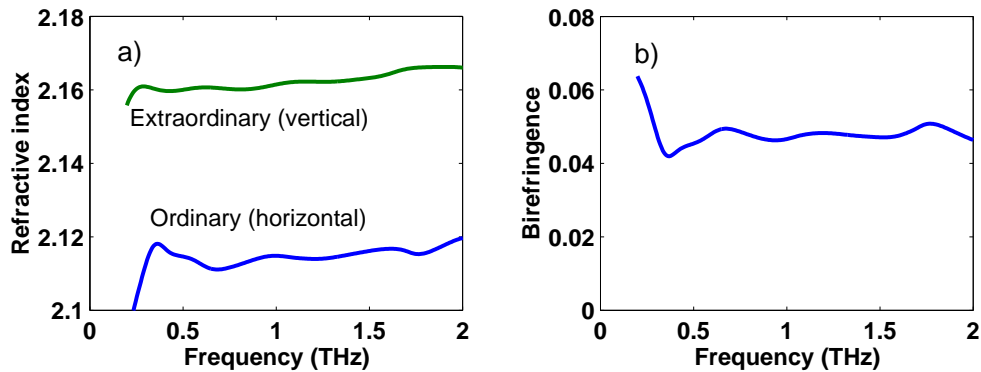}
    \caption{\label{FIG:ribi}
    a) Ordinary and extraordinary refractive indices of quartz measured by polarisation sensitive THz-TDS.
    b) Birefringence of quartz calculated form the refractive indices.
    }
\end{figure}

In order to obtain the refractive indices it is necessary to do an
analysis by frequency components. By applying a Fourier transform to
${\bf{E}} _{\rm ref}(t)$ and ${\bf{E}}_{\rm sam}(t)$ the vectors
${\bf{E}}_{\rm ref}(\omega)$ and ${\bf{E}}_{\rm sam}(\omega)$ are obtained
which are in general complex, the components of these vectors can be
expressed in the form
\begin{equation}
{E_{\rm ref}^{[k]}(\omega)=\mathcal{E}_{\rm ref}^{[k]}(\omega)\,e^{i\phi_{\rm ref}^{[k]}(\omega)}
 }\label{e:componentsr}
\end{equation}
and
\begin{equation}
{E_{\rm sam}^{[k]}(\omega)=\mathcal{E}_{\rm sam}^{[k]}(\omega)\,e^{i\phi_{\rm sam}^{[k]}(\omega)}
 }\label{e:componentss}
\end{equation}
where $\mathcal{E}$ and $\phi$ are the complex amplitude and phase
of the respective vector component. The refractive index for
component $k$ will be given by
\begin{equation}
{n_k=1+c\frac{\phi_{\rm sam}^{[k]}(\omega)-\phi_{\rm ref}^{[k]}(\omega)}{\omega
d},
 }\label{e:refindex}
\end{equation}
here $d$ is the thickness of the sample and $c$ the speed of light.
Notice that no assumption has been made on which components of the
electric field the index $k$ refers to, these components could be
written in the plane polarised base but in general could be any
elliptical components (including the right and left circular
components). Given that this experiment was planned so that the
receiver and crystal axes coincide, the appropriate base to use in
this case is the plane one in the horizontal and vertical directions
that are parallel to the Y (normal) and Z (extraordinary) axes of
the quartz sample. Substituting $E_{\rm horizontal}$ and $E_{\rm
vertical}$ as measured in Equation~\ref{e:refindex}, the two
refractive indices are obtained, as shown in Figure~\ref{FIG:ribi}a.
Both curves of Figure~\ref{FIG:ribi}a match very well with values
already measured separately by time domain spectroscopy for the
ordinary and extraordinary refraction
indices.\cite{grischkowsky:2006} The main advantage of this
technique is that it is based on detecting the full polarisation of
the initial and final pulses, allowing the study of not just
linearly polarised states but any arbitrary changes in the state of
polarisation. Finally we obtain the difference of the extraordinary
and the ordinary refractive indices which is shown in
Figure~\ref{FIG:ribi}b.

\section{Conclusion}
\label{s:conclusion}

The process of design, construction and characterisation of a
polarisation sensitive photoconductive receiver for terahertz time
domain spectroscopy was presented. This detector extends the power
of traditional THz-TDS by measuring both transverse components of
the electric field of a THz wave simultaneously. This allows
broad-band studies in the far infrared of polarisation dependent
complex properties of materials. We have demonstrated this by
measuring the ordinary and extraordinary refractive indices of
quartz in the range 0.2 to 2.0\,THz. This detector is expected to
have many applications in the near future, in particular it will be
fundamental for developing a terahertz time domain circular
dichroism spectrometer.

\acknowledgments     
The authors would like to thank the EPSRC and the Royal Society (UK) for financial
support of this work. E.C.C.\ wishes to thank CONACyT (M$\acute{\rm{e}}$xico) for
financial support.


\bibliographystyle{spiebib}              

\end{document}